\documentclass[aps,prc,twocolumn,floatfix]{revtex4}
\usepackage{graphicx,amsmath,amssymb}
\usepackage{bm}
\usepackage{longtable}

\begin{document}

\title{Shell Model Monte Carlo method in the $pn$-formalism \\  and 
       applications to the Zr and Mo isotopes}

\author{C. \"Ozen$^1$ and D.J. Dean$^2$}
\affiliation{$^1$Department of Physics and Astronomy, 
University of Tennessee, Knoxville, Tennessee 37996}
\affiliation{$^2$Physics Division, Oak Ridge National 
Laboratory, Oak Ridge, Tennessee 37831}

\date{\today}

\begin{abstract}
We report on the development of a new shell-model Monte Carlo algorithm which  
uses the proton-neutron formalism. Shell model Monte Carlo 
methods, within the isospin formulation,  have been successfully 
used in large-scale shell-model calculations.
Motivation for this work is to extend the feasibility of these methods 
to shell-model studies involving non-identical 
proton and neutron valence spaces.
We show the viability of the new approach with some 
test results. Finally, we use a 
realistic nucleon-nucleon 
interaction in the model space described by 
$(1p_{1/2},0g_{9/2})$ proton 
and $(1d_{5/2},2s_{1/2},1d_{3/2},0g_{7/2},0h_{11/2})$ neutron 
orbitals above the 
${}^{88}\mathrm{Sr}$ core to  calculate ground-state energies, 
binding energies, $B(E2)$ strengths,  and to study pairing 
properties of 
the even-even ${}^{90-104}\mathrm{Zr}$ and ${}^{92-106}\mathrm{Mo}$ 
isotope chains.
\end{abstract}

\pacs{}

\maketitle

\section{\label{sec:intro}Introduction and Motivation}

The Shell model Monte Carlo (SMMC) 
method~\cite{koo1,koo2,lang1} was developed as an
alternative to direct diagonalization in order
to study low-energy nuclear properties.  It was successfully 
applied to nuclear problems where large model spaces
made diagonalization impractical. 
In the canonical SMMC approach, one calculates 
the thermal expectation values 
of observables of few-body operators 
by representing the imaginary-time many-body 
evolution operator as a superposition
of one-body propagators in fluctuating auxiliary fields.
Thus, one recasts the Hamiltonian diagonalization 
problem as a stochastic integration problem. 

In this paper, we report on the development of an SMMC approach in
the $pn$-formalism. 
This implementation of SMMC enables one to treat 
shell-model Hamiltonians that are not isospin invariant 
in the model space, or for which different model spaces are 
used for protons and neutrons.  In the following, we will 
use the abbreviated form SMMCpn, to distinguish the approach
discussed here from the original one. We note that special features
of the pairing+quadrupole interaction enabled a special implementation
of SMMC in non-degenerate proton and neutron model spaces for
calculations in rare earth nuclei \cite{dean93, white00}. The method
presented in this work is general and may be used 
for realistic Hamiltonians, as well those of a more schematic variety. 

As a first novel application of the new implementation, we perform
shell-model calculations for 
the even-even  ${}^{90-104}\mathrm{Zr}$ and ${}^{92-106}\mathrm{Mo}$ 
isotopic chains. Initial experimental studies \cite{exp70} indicated
that nuclei in this region have very large deformations, and that 
the transition from spherical shapes to highly deformed shapes 
occurs abruptly: $^{96}$Zr is rather spherical, while $^{100-104}$Zr
nuclei are well deformed with a quadrupole deformation parameter
of $\beta_2=0.35$ \cite{exp90}. Furthermore, the spherical-to-deformed
transition is more abrupt in the Zr isotopes than in the 
nearby elements Mo, Ru, and Pd. Generator-coordinate mean-field calculations 
in this region \cite{skalski93} are able to reproduce the shape
transitions with particular Skyrme interactions. Furthermore, 
the region exhibits significant shape-coexistence phenomena 
\cite{reinhard99,wood92}.

The history of shell-model applications in
this mass region goes back to the 1960s with model spaces built 
on ${}^{88}\mathrm{Sr}$ or 
${}^{90}\mathrm{Zr}$ cores~\cite{tal1,aur1,ver1,coh1}. 
Gloeckner~\cite{glo1} used an
effective interaction built on a ${}^{88}\mathrm{Sr}$ core 
with a model space
consisting of the orbitals $\pi:(1p_{1/2},0g_{9/2})$, 
$\nu:(1d_{5/2},2s_{1/2})$. Other studies 
used  larger model spaces~\cite{ips1,hals1,zhang1} with 
varying effective interactions and 
truncation schemes.  
Holt {\it et. al.}~\cite{holt1} derived a realistic effective interaction 
using many-body perturbation techniques \cite{hj95}
in the model space 
$\pi:(1p_{1/2},0g_{9/2})$, $\nu:(1d_{5/2},
2s_{1/2},1d_{3/2},0g_{7/2},0h_{11/2})$. This effective interaction
was based on the realistic nucleon-nucleon CD-Bonn 
potential \cite{mach96}, and shell-model 
diagonalization calculations were carried out for the 
low-lying spectra of the Zr isotopes with neutron numbers 
from $N=52$ to $N=60$. Their results showed reasonable agreement with 
experimental spectra. In this article, we use a 
slightly modified version of this realistic effective interaction to 
explore Zr and Mo nuclei through $N=64$. 

In Sec.~\ref{sec:form}, we give an outline of the SMMC method 
with an emphasis on the
differences in the SMMCpn implementation when compared to the 
isospin-conserving implementation.
Then, in Sec.~\ref{subsec:comp},
we will  demonstrate the utility of the new approach by a comparison of various
numerical results obtained using the SMMCpn technique
for a few $fp$-shell nuclei
to those calculated by direct diagonalization and earlier SMMC studies.
Calculations for the Zr and Mo isotopes, which are 
presented in Sec.~\ref{subsec:appl},
were carried out in the same model space 
as in  Holt {\it et. al.}~\cite{holt1},
using a slightly modified interaction \cite{juodagalvis1}.
We show results for ground-state energies, binding energies, $B(E2)$ 
strengths, and BCS-like pairing correlations for the Zr and Mo isotope chains. 
We conclude with a perspective on this avenue of research.

\section{\label{sec:form}Formalism}

In the SMMC method, we calculate 
expectation values of operators within a thermal
ensemble of particles whose interactions are governed
by the Hamiltonian ${\hat H}$ of the system. 
(A zero-temperature formalism also exists but will not be discussed here.) 
The canonical expectation value of an
operator $\hat{X}$ at a temperature $T$ is given by
\begin{equation}
\langle \hat{X} \rangle = \frac{{\rm Tr}[\hat{U}\hat{X}]}{Z}\;, 
\end{equation}
where the partition function of the system is given 
by $Z(\beta)={\rm Tr}{\hat U  }$,
$\beta=1/{T}$ is the inverse temperature (with units MeV$^{-1}$),
and the many-body evolution operator is
$\hat{U}=e^{-\beta\hat{H}}$. 
The quantum-mechanical trace of an operator is defined as
\begin{equation}
{\rm Tr}{\hat X} = \sum_\alpha \langle\alpha\mid\hat{X}\mid\alpha\rangle\;,
\end{equation}
where the sum runs over {\it all} many-body states in the Hilbert space. 
For nuclear calculations, the number of valence particles is usually limited,
so that number projection becomes important. The original SMMC 
method preserved isospin within the same neutron and proton 
model spaces.  We discuss in the following
how to implement number projection when the isospin quantum
number is broken.  Note that in the limit of $\beta \rightarrow \infty$, we
may evaluate ground-state properties of the nuclear systems. 

In the following, we will consider Hamiltonians that have at most 
two-body terms. Any such Hamiltonian can be cast into a quadratic form:

  \begin{eqnarray}
     \hat{H}=\sum_i \varepsilon_\alpha {\hat\rho}_\alpha +
             \frac{1}{2} \sum_{\alpha} \lambda_{\alpha}
             {\hat{\rho}}_{\alpha}^{2},
     \label{eq:quadHam}
  \end{eqnarray}

\noindent where $\varepsilon_\alpha$ is the  energy of 
single-particle level $\alpha$, and
the operator $\hat{\rho}_{\alpha}$ is a a one-body 
density operator of the form $a^{\dagger} a $. Details are
given in \cite{koo1} on 
how to transform one-body operators with 
quantum numbers $\{n,l,j,j_z,t_z\}$ (where $n$ is the 
principal quantum number, $l$ is the orbital momentum, $j$ is the 
total angular momentum, and $j_z$ is its projection, and $t_z=\pm 1$ 
for protons and neutrons) to 
the form shown in Eq.~(\ref{eq:quadHam}).

At the heart of the SMMC method lies the 
linearization of the imaginary-time many-body propagator. 
Since, in general, $[\hat {\rho}_\alpha, \hat{\rho}_\beta]\not=0$,
we must split the interval $\beta$ into $N_t$ ``time slices'' of length
$\Delta\beta\equiv\beta/N_t$.
We apply the Hubbard-Stratonovich 
transformation~\cite{hub1,str1} to the two-body evolution operator at
each time slice. 
In compact notation, the partition function can 
be written as:
\begin{eqnarray}
Z(\beta) & = & {\rm Tr}\hat{U}={\rm Tr}e^{-\beta {\hat H}}\longrightarrow{\rm Tr}\left[e^{-\Delta\beta
{\hat H}}\right]^{N_t} \nonumber \\ 
& \longrightarrow  &\int {\mathfrak D}[\sigma] 
G(\sigma) {\rm Tr} \hat{U}_\sigma\;,
\end{eqnarray}
where the metric of the functional integral is
\begin{eqnarray}
\mathcal{D}[\sigma]=\prod_{\alpha,n} \sqrt{\frac{\Delta\beta \vert V_\alpha \vert}{2\pi}}
\mathrm{d} \sigma_\alpha (\tau_n)\;,
\end{eqnarray}
and the Gaussian weight is given by
  \begin{eqnarray}
     G_\sigma=e^{ -\frac{1}{2}\Delta\beta\sum_{\alpha,n}
              \vert \lambda_\alpha \vert \sigma_\alpha^2(\tau_n)}\;. 
  \end{eqnarray}
The one-body evolution operator is written as 
\begin{eqnarray}
\hat{U}_\sigma = 
\prod_{n=1}^{N_t} e^{-\Delta\beta {\hat h}_\sigma(\tau_n)} 
\equiv \mathcal{T} e^{-\int_0^\beta
\! \mathrm{d} \tau
\hat{h}_\sigma (\tau)}\;, 
\end{eqnarray}
where we note the dependence on the auxiliary 
fields $\sigma(\tau_n)$. This time-ordered product means that this 
formalism yields a path integral in the fields $\sigma$. 
The linearized one-body Hamiltonian for 
the time slice $\tau_n$ is given by
  \begin{eqnarray}
     \hat{h}_\sigma(\tau_n)=
     \sum_i \varepsilon_\alpha \hat{\rho}_\alpha +\sum_{\alpha n} 
     s_\alpha \lambda_\alpha \sigma_\alpha (\tau_n)
     \hat{\rho}_\alpha,
     \label{eq:linham1}
  \end{eqnarray}
with $s= \pm 1$, if $ \lambda > 0$; or $ s= \pm i $, if $\lambda <0$.
Note that because the 
various $\hat {\rho}_\alpha$ need not commute, (3.18) is
accurate only through order $\Delta\beta$ and that
the representation of $e^{-\Delta\beta\hat h}$ must 
be accurate through order $\Delta\beta^2$ to achieve that accuracy.

The thermal expectation values can be expressed as the ratio of
path integrals in 
fluctuating auxiliary fields,

  \begin{eqnarray}
     \langle \hat{O} \rangle = \frac{\mathrm{Tr}[\hat{O}e^{-\beta \hat{H}}]}
                                    {\mathrm{Tr}[e^{-\beta \hat{H}}]}
                                    =\frac{\int \! \mathcal{D}[\sigma] G_\sigma
                                     \langle \hat{O} \rangle_\sigma
                                     \xi_\sigma}
                                     {{\int \! \mathcal{D}[\sigma] G_\sigma
                                     \xi_\sigma }}
     \label{MultInt1}
  \end{eqnarray}

 \noindent  where the following definitions are used:

  \begin{eqnarray}
      \xi_\sigma=\mathrm{Tr} \hat{U}_\sigma , \qquad
         \langle \hat{O} \rangle_\sigma = \frac{\mathrm{Tr}[\hat{O} \hat{U}_\sigma]}
         {\mathrm{Tr} \hat{U}_\sigma}.
      \label{eq:traces}
  \end{eqnarray}

In order to use Metropolis Monte Carlo 
sampling methods \cite{metrop}, we need to 
define a positive-definite weight function,

  \begin{eqnarray}
     \label{eq:W}
     W_\sigma = G_\sigma \vert \xi_\sigma \vert,
  \end{eqnarray}

  \noindent  so Eqn.~\ref{MultInt1} can now be rewritten as

  \begin{eqnarray}
      \langle \hat{O} \rangle = \frac{\int \! \mathcal{D}[\sigma] W_\sigma
                                      \langle \hat{O} \rangle_\sigma
                                      \Phi_\sigma}
                                     { \int \! \mathcal{D}[\sigma] W_\sigma \Phi_\sigma}
                              \equiv \frac{\langle \langle \hat{O} \rangle_\sigma
                                     \Phi_\sigma  \rangle_W}
                                     {\langle \Phi_\sigma  \rangle_W}
      \label{MultInt2}
  \end{eqnarray}

  \noindent where

  \begin{eqnarray}
     \Phi_\sigma = \frac{\xi_\sigma}{\vert \xi_\sigma \vert}
  \end{eqnarray}

  \noindent is the sign of the partition function.

The description above shows how one may transform the shell model into 
a problem of quadrature integration. 
Objects, $\xi_\sigma$ and 
$\langle \hat{X} \rangle_\sigma$, in the 
integrands are of one-body nature and 
are represented by 
$N_s \times N_s$  dimensional matrices where $N_s$ is 
the number of the single-particle levels
in the valence space. 
The path integrals in the auxiliary fields are evaluated by performing 
a Metropolis random walk in the field space. Thermodynamic expectation 
values are given as the ratio
of two multidimensional integrals over the auxiliary fields. 
The dimension $D$ of these integrals is of 
order $N_s^2N_t$, which can exceed $10^5$ for
the problems of interest in this paper.

Note that the Monte Carlo 
sign problem enters calculations when any 
of the $\lambda_\alpha$ matrix elements
is positive. Realistic shell-model interactions always
have such terms; a special case is the pairing-plus-quadrupole Hamiltonian
which has no sign problems. 

If the proton and neutron valence spaces are identical
and the Hamiltonian $\hat{H}$ 
is  isospin-symmetric, then the Hamiltonian can be cast into a quadratic
form which respects this symmetry explicitly.
In that case, it is possible to form linear combinations 
of density operators that
separately conserve the neutron and proton numbers. 
In the isospin formulation 
(as done in the original SMMC
studies) proton and neutron many-body wave functions are
represented by Slater determinants, 
and the ensuing one-body propagator factors into two propagators as well,
one for protons and another for neutrons. 
The canonical traces are then calculated by 
applying the number-projection operator 
to obtain the
desired proton and neutron numbers. In contrast, we will
employ the shell-model 
Monte Carlo method to Hamiltonians that are 
not necessarily isospin invariant or for which 
proton and neutron valence spaces are different. 
A relaxation of the isospin symmetry in the quadratic 
forms becomes essential in order to
employ the SMMC method in such cases. 
In the $pn$-formalism,
proton and neutron valence spaces are no longer 
distinguished from each other; instead,
we consider a single valence space containing both 
proton and neutron orbitals. 
In this way, the
density operators in the one-body Hamiltonian $h(\sigma)$
inevitably 
mix protons and neutrons, and
as a consequence their respective expectation values will 
fluctuate from sample to sample.  
The canonical trace is then retrieved by 
employing projection operators to fix the total number of 
particles $A$ and the $z$-component of the total isospin $T_z$. 
This implementation represents the major difference between 
the original SMMC and the SMMCpn techniques. 

Projection operators for fixed $A$ and $T_z$ are given by 

\begin{subequations}
  \begin{eqnarray} 
    \hat P_A =
    \int^{2\pi}_0 {\frac{d\phi}{2\pi}} e^{-i\phi A}
    e^{i\phi\hat N}  
  \end{eqnarray}

  \noindent  and

   \begin{eqnarray}
    \hat  P_{T_z} &=& \int^{2\pi}_0 {\frac{d\theta}{2\pi}} e^{-i\theta T_z}
    e^{i\theta \hat T_z}.   
   \end{eqnarray}
\end{subequations}

\noindent respectively. In the discrete Fourier 
representation, we make the substitution,

 \begin{eqnarray}
   \int^{2\pi}_0 {\frac{d\phi}{2\pi}} \int^{2\pi}_0  {\frac{d\theta}{2\pi}} \longrightarrow
   \frac{1}{N_s (N_{T_z}+1)} \sum_{m=1}^{N_s} \sum_{n=0}^{N_{T_z}},
 \end{eqnarray}

 \noindent  where the quadrature points are given by $\phi_m=2\pi m/N_s$ and 
 $\theta_n=2\pi n/(N_{T_z}+1)$,
and
$N_{T_z}$ is the number of values
 $T_z$ can take.
 As an example, the canonical trace of the one-body 
propagator $\hat{U}_\sigma$ can be obtained 
 by acting with both  projection operators on the 
grand-canonical trace,
 ${\rm Tr}\hat U_\sigma=\det (1+{\bf U_\sigma})$:

 \begin{eqnarray}{\rm Tr}_{A,T_z} \hat U_\sigma &=&
  \frac{1}{N_s (N_{T_z}+1)} \sum_{m,n} 
  e^{-i\phi A} e^{-i\theta T_z} \nonumber \\   
  & \times&   {\rm Tr}\,
  e^{i\phi \hat N} e^{i\theta\hat T_z}  \hat U_\sigma \\
  &=& \frac{1}{N_s (N_{T_z}+1)} \sum_{m,n} 
  e^{-i\phi A} e^{-i\theta T_z}  \times \nonumber \\
  &\times  &\det (1+e^{i\phi}
  e^{i\theta {\bf T_z}} {\bf U_\sigma} ) \;,
 \end{eqnarray}

 \noindent where  the boldface symbols are used for the matrix 
representation of the operators,
 for example

 \begin{eqnarray}
 {\bf T_z}= \frac{1}{2}
 \left(\begin{array}{c|c}
 {\, -\bf 1} & {\bf \; 0} \\
 \hline
 {\bf \; 0} & {\bf \; 1}
 \end{array}\right).
 \end{eqnarray}

 A typical difficulty in the SMMC applications is due to a sign problem
 arising from the repulsive part of the realistic interactions.
In the case of realistic interactions, a straightforward
application of Eqn.~(\ref{eq:W}) to obtain a positive definite 
weight will introduce a highly fluctuating integral. This will 
give rise to expectation values with very large fluctuations.  In order
to avoid this situation, 
 we adopt a practical solution~\cite{alha1} to the sign  problem by 
 breaking the two-body interaction into
``good" (without a sign problem) and ``bad" (with a sign problem) parts: 
$H=H_{G}+H_{B}$.  Using a parameter $g$, we then construct a new family of 
Hamiltonians $H(g)=H_{G}+gH_{B}$  which are free of the 
sign problem for non-positive values of $g$.
The SMMC observables are evaluated for a number 
of different $g$ values in the 
interval $-1 \leq g \leq 0$, and the 
physical values are thus retrieved 
by extrapolation to $g=1$. We use polynomial extrapolations
and choose the minimum order that 
gives $\chi \approx 1$. In our calculations, most of
the extrapolations are linear or quadratic.
In the extrapolation of $\langle H \rangle $, the variational principle 
imposes a vanishing derivative
at the physical value $g=1$.  
A cubic extrapolation in this case
typically gives the best results.

Another problem encountered in applying the SMMC methods concerns efficiency
of the Metropolis algorithm in generating uncorrelated field configurations.
Rather than sample continuous fields, where decorrelated samples are obtained
after only very many (of order 200) Metropolis steps, we 
approximate the continuous integral over each
$\sigma_\alpha(\tau_n)$ by a discrete sum derived from a Gaussian quadrature
\cite{dean93}.
In particular, the relation
\begin{equation}
e^{\Delta\beta \lambda \hat {\rho}^2/2}\approx
\int^\infty_{-\infty} d\sigma f(\sigma)
e^{\Delta\beta \lambda \sigma \hat{\rho}}
\end{equation}
is satisfied through terms in $(\Delta\beta)^2$ if
\begin{equation}
f(\sigma)=\frac{1}{6}
\left[\delta(\sigma-\sigma_0)+\delta(\sigma+\sigma_0)+
4\delta(\sigma)\right]\;,\end{equation}
where {$\sigma_0=(3/\lambda\Delta\beta)^{1/2}$}. 
In the SMMCpn algorithm, we find that samples are well 
decorrelated after only a few (typically less than ten) Metropolis steps 
using these descretized fields. 

We describe in the following section our initial results using the 
SMMCpn method. 

\section{\label{sec:res}Results}

\subsection{Comparisons with direct diagonalization 
and previous SMMC calculations}
\label{subsec:comp}

In this section, we present a number of test cases that validate the
SMMCpn approach.  For this purpose, we first 
carried out calculations on a few $sd$-shell nuclei using a 
quadrupole-plus-pairing interaction which is free of the sign-problem. 
This interaction can be written as 

 \begin{eqnarray}
   \hat{V}=- \chi \hat{Q} \cdot \hat{Q}-g \hat{P}^{(0,1)\dagger} \cdot {\tilde{\hat{P}}{}^{(0,1)}}
 \end{eqnarray} 

\noindent where  

 \begin{eqnarray}
     \hat{Q} = \frac{1}{\sqrt{5}}\sum_{a b}
     \langle j_a \vert\vert \frac{dV_{WS}}{dr}Y_2 \vert\vert j_b \rangle
     [a_{j_a}^{\dagger} \otimes \tilde{a}_{j_b}]^{J=1,T=0}
 \end{eqnarray}

 \noindent and 

 \begin{eqnarray}
     & &\hat{P}^{(0,1)\dagger} = \sum_{a} [a_{j_a}^{\dagger} \otimes a_{j_a}^{\dagger}]^{J=0,T=1}.
 \end{eqnarray}

 \noindent The strength of the interaction terms were chosen as 
 $\chi=0.0260$ $\mathrm{Me{V}^{-1}fm^2}$ and 
 $g=0.212$ MeV. We adopted the standard 
USD \cite{wild84} single-particle energies.
 The term $V_{WS}$ in the equation above is the central part of a 
 Woods-Saxon potential with parameters given in~\cite{Boh1}.
 The SMMCpn calculations were performed at 
 $\beta=2$ $\mathrm{MeV}^{-1}$ with $N_t=128$ time 
slices ($\Delta\beta=1/64$~MeV$^{-1}$), and each calculation involved 
2500-3000 uncorrelated samples. 
Note that typical isospin-conserving SMMC calculations only require 
$\Delta\beta=1/32$~MeV$^{-1}$ for similar convergence. 
 A comparison of our results with the direct diagonalization 
 values (obtained by running the ANTOINE~\cite{antoi} code)
 is given in Table~\ref{Tab:sd_ene}. Results are compatible within the 
 internal heating energy and statistical errors 
of the thermal SMMC calculations.

\begingroup
\squeezetable
\begin{table}[!hpt]
\caption
        {Comparison of the ground-state 
energies (in MeV), as calculated by SMMCpn and the
ANTOINE results. A quadrupole-plus-pairing interaction is used.}

\label{Tab:sd_ene}
\begin{center}
\begin{tabular}{ccc}
\hline
\hline
Nucleus &   E (ANTOINE)     &   E (SMMCpn) \\
\hline
\hline
${}^{24}\mathrm{Mg}$   & $-39.28$ &  $-38.68\pm0.27$ \\
${}^{26}\mathrm{Mg}$   & $-43.58$ &  $-42.75\pm0.70$ \\
${}^{22}\mathrm{Ne}$   & $-30.23$ &  $-29.53\pm0.41$ \\
${}^{28}\mathrm{Si}$   & $-49.25$ &  $-48.80\pm0.37$ \\
\hline
\end{tabular}
\end{center}
 \vspace{1.0cm}
\end{table}
\endgroup

 We also
 tested the viability of the new implementation with the utilization of
 the extrapolation method described above. 
For this purpose, a few  $fp$-shell nuclei 
 were chosen and  the 
 modified Kuo-Brown KB3 residual interaction \cite{kb3} was used. 
 We calculated the ground-state energies,  
 total $B(M1)$,  $B(E2)$, and the Gamow-Teller strengths
 of the nuclei and compared our results to those obtained by
 exact diagonalization \cite{cau1} and those obtained by
 the isospin SMMC \cite{la1}.
 The SMMCpn calculations were
 performed at $\beta=2$ $\mathrm{ MeV}^{-1}$ with $\Delta\beta=1/64$MeV$^{-1}$,
and each calculation at each of six values of the extrapolation parameter $g$
involved 8000-9000 uncorrelated samples. 
We used a quadratic extrapolation for the  total $B(M1)$ and $B(GT+)$ strengths, while
for the total $B(E2)$ strengths, a linear extrapolation 
was more reasonable. The ground state energies
employed  cubic polynomials subject to the 
constraint $d\langle H \rangle / dg \vert_{g=1}=0$ due 
to a variational principle 
that $\langle H \rangle$ obeys. In all the cases, the errors were 
conservatively adopted from a quadratic extrapolation.  

\begingroup
\squeezetable
\begin{table}[!hpt]
\caption
        {Comparison of exact diagonalization, 
  SMMCpn, and isospin SMMC results for
 valence space energies 
 (in $\mathrm{MeV}$) and $B(E2)$ strengths (in $\mathrm{e^2 fm^4}$).
 Typical error bar for energies is 
 $\pm0.6$ MeV for SMMCpn and $\pm0.4$ MeV for isospin SMMC
 calculations.}
\label{Tab:e2etc}
\begin{center}
\begin{tabular}{cccc}
\hline
\hline
Nucleus &   E     &   E       &     E          \\
        &  exact  & SMMC(pn)  &  SMMC (iso)    \\
\hline
${}^{48}\mathrm{Ti}$   & $-24.6$ &  $-24.4$  & $-23.9$  \\
${}^{48}\mathrm{Cr}$   & $-32.9$ &  $-32.6$  & $-32.3$  \\
${}^{56}\mathrm{Fe}$   & $-66.4$ &  $-66.0$  & $-65.8$  \\
${}^{64}\mathrm{Zn}$   & $-106.3$ & $-106.5$ & $-104.8$ \\
\hline
\vspace{0.5cm} \\
\hline
\hline
Nucleus & $\sum$B(E2)  & $\sum$B(E2) & $\sum$ B(E2)  \\
        &   exact      &    SMMC(pn) &  SMMC (iso)   \\
\hline
${}^{48}\mathrm{Ti}$   & $476$   &  $459\pm33$     & $455\pm25$    \\
${}^{48}\mathrm{Cr}$   & $978$   &  $745\pm40$     & $945\pm45$    \\
${}^{56}\mathrm{Fe}$   & $1019$  &  $913\pm55$     & $990\pm6\phantom{1}$     \\
${}^{64}\mathrm{Zn}$   & $1157$  & $1116\pm81$ & $1225\pm65$ \\
\hline
\end{tabular}
\end{center}
 \vspace{1.0cm}
\end{table}
\endgroup

 In Table~\ref{Tab:e2etc}, ground-state energies and 
$B(E2)$ strengths are tabulated.
 In all cases, the energies agree strikingly 
well within  error bars that are reasonable
 with the internal excitation energy of a 
few hundred keV due to the finite-temperature
 calculations. The $B(E2)$ strength is given by
\begin{eqnarray}
  B(E2)=\langle (e_p \hat{Q}_p + e_n \hat{Q}_n)^2 \rangle,
\end{eqnarray}
where the quadrupole operator is defined as
 $\hat{Q}_{p(n)}=\sum_i r_i^2 Y_2(\theta_i,\phi_i)$. The effective charges were chosen to be
 $e_p=1.35$ and $e_n=0.35$, and the oscillator strength is given by  $b=1.01A^{1/6}$.
 $B(E2)$ values are also nicely reproduced in 
general, while in the case of  ${}^{48}\mathrm{Cr}$,
 the exact result is underestimated by $\approx$ 25\%.

\begingroup
\squeezetable
\begin{table}[!t]
\caption 
    {Comparison of exact diagonalization, SMMCpn, and isospin SMMC results for
    $B(M1)$ (in $\mu_N$) and Gamow-Teller strengths.}
\label{Tab:m1etc}
\begin{center}
\begin{tabular}{cccc}
\hline
\hline
Nucleus & $\sum$B(M1) & $\sum$B(M1) & $\sum$B(M1)  \\
        &  exact  & SMMC(pn)  &  SMMC (iso)        \\
\hline
\hline
${}^{48}\mathrm{Ti}$   & $10.6$  &  $10.4\pm4$   & $10.2\pm1.2$    \\
${}^{48}\mathrm{Cr}$   & $12.0$  &  $12.5\pm4.4$   & $13.8\pm1.7$  \\
${}^{56}\mathrm{Fe}$   & $19.4$  &  $22.6\pm6.4$   & $20.4\pm3.0$  \\
${}^{64}\mathrm{Zn}$   & $21.6$  &  $22.8\pm1.2$   & $23.6\pm2.2$  \\
\hline
\vspace{0.5cm} \\
\hline
\hline
Nucleus  & $\sum$B(GT+) & $\sum$B(GT+) & $\sum$B(GT+)  \\
         &  exact       &  SMMC(pn)   & SMMC (iso)     \\

\hline
${}^{48}\mathrm{Ti}$   & $1.26$  &  $0.89\pm0.36$  & $1.13\pm0.18$ \\
${}^{48}\mathrm{Cr}$   & $4.13$  &  $4.35\pm0.44$  & $4.37\pm0.35$ \\
${}^{56}\mathrm{Fe}$   & $4.69$  &  $4.02\pm0.55$  & $3.99\pm0.27$ \\
${}^{64}\mathrm{Zn}$   & $5.54$  &  $5.66\pm0.7$   & $4.13\pm0.34$ \\
\hline
\end{tabular}
\end{center}
 \vspace{1.0cm}
\end{table}
\endgroup

 Table~\ref{Tab:m1etc} shows 
a comparison of results for the $B(M1)$ and $B(GT+)$ strengths.
 The $B(M1)$ strength is defined by
\begin{eqnarray}
B(M1)=\langle (\sum_i \mu_N \{ g_l \vec{l}+g_s \vec{s} \})^2\rangle
\end{eqnarray}
where $\mu_N$ is the nuclear magneton. We used the bare g factors
 for angular momentum and spin
 ($ g_l=1$ and $g_s=5.586$ for protons and $g_l=0$ and $g_s=-3.826$ for neutrons).
 The Gamow-Teller strength is defined by
 \begin{eqnarray}
    B(GT+)=\langle G_{\mp}G_{\pm} \rangle,
 \end{eqnarray}
where the unquenched Gamow-Teller operator is 
written as $G_{\pm}=\sum_i \vec{\sigma} t_{\pm}$.
 Both $ B(M1)$ and $B(GT+)$ results agree well with those obtained by direct diagonalization. Lastly,
 as in the case of isospin SMMC calculations, 
our calculations satisfy the Ikeda sum rule $B(GT-)-B(GT+)=3(N-Z)$ exactly.

\subsection{Applications to Zr and Mo isotopes}
\label{subsec:appl}
Calculations for the Zr and Mo isotopes were 
carried out in  the same valence space as in 
Holt {\it et. al.}~\cite{holt1},
which is built on the ${}^{88}\mathrm{Sr}$ core,
but we employed a slightly modified interaction which has been
previously tested for nuclei with small numbers of valence 
particles in this region \cite{juodagalvis1}.
The model space and single-particle energies that were used 
in the calculations are given in 
Table~\ref{tabl:physparm} (\cite{holt1} and references therein). 
All calculations
were performed at $\beta=2$ $\mathrm{MeV^{-1}}$ using $N_t=128$ time intervals
with 8500-9500 uncorrelated samples for each extrapolation parameter $g$.

\begingroup
\squeezetable
\begin{table}[!hpt]
\caption{The model space and single-particle energies used in these 
calculations.}
\label{tabl:physparm}
\begin{center}
\begin{tabular}{|lllll|}
 \hline
 \multicolumn{2}{|c}{Protons} &\multicolumn{1}{c}{} &  \multicolumn{2}{c|}{Neutrons}  \\
 \hline
 \hline
 Orbital &  Energy (MeV) &  & Orbital & Energy (MeV) \\
 \hline
 \hline
   $0\mathrm{g}_{9/2}$  & 0.90 & & $0\mathrm{h}_{11/2}$ & 3.50  \\
   $1\mathrm{p}_{1/2}$  & 0.00 & & $0\mathrm{g}_{7/2}$  & 2.63  \\
                        &      & & $1\mathrm{d}_{3/2}$  & 2.23  \\
                        &      & & $2\mathrm{s}_{1/2}$  & 1.26  \\
                        &      & & $1\mathrm{d}_{5/2}$  & 0.00  \\
                        &      & &                      &       \\
 \multicolumn{2}{|c}{$e_p=1.8$} & &  \multicolumn{2}{c|}{$e_n=1.5$} \\
                      & &     $b=2.25 \; \mathrm{fm}$    &  &    \\
 \hline
\end{tabular}
\end{center}
 \vspace{1.0cm}
\end{table}
\endgroup

\subsubsection{Ground-state energies}

\begin{figure}
   \rotatebox{0}{\scalebox{0.62}{\includegraphics{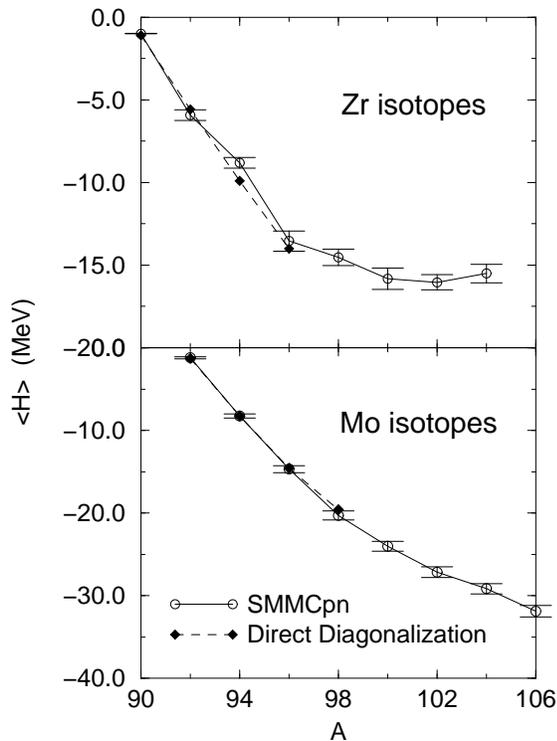}}}
 \caption{\label{fig:ene} Ground-state energies of Zr and Mo
          isotopes. Results from direct diagonalization available for 
lighter isotopes are also shown.}
 \end{figure}

Shown in Fig.~\ref{fig:ene} is the comparison 
of the expectation value of energy
$\langle \hat{H} \rangle$. Filled circles
that are connected with dashed 
lines represent exact
diagonalization results \cite{Juo1} and in 
both the Zr and Mo cases,
the agreement of the SMMCpn values is remarkable. 
Only for ${}^{94}\mathrm{Zr}$ do error bars of the 
SMMCpn result miss the exact diagonalization value slightly. 
This represents the full test of the SMMCpn approach. 

\subsubsection{Binding energies}
The ground-state energies shown in Fig.~\ref{fig:ene} correspond to
the contribution to the nuclear binding energy of the interaction 
of the valence particles among themselves.

\begin{figure}
   \rotatebox{0}{\scalebox{0.62}{\includegraphics{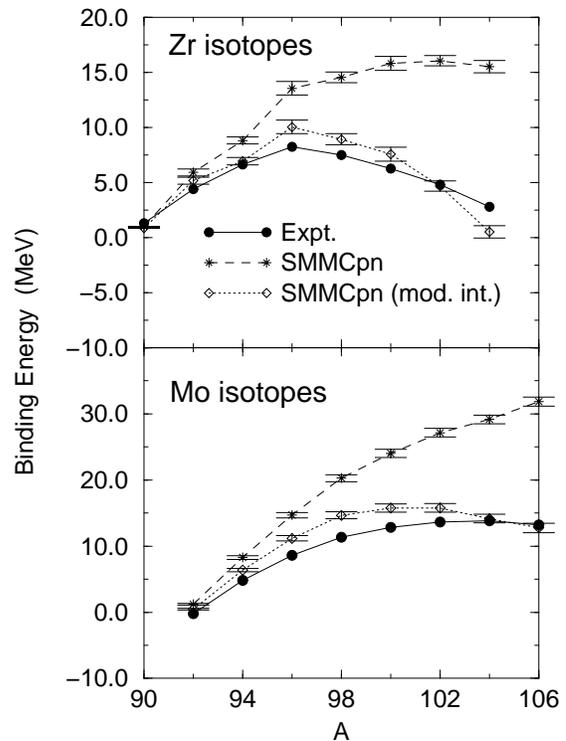}}}
 \caption{\label{fig:bind} Binding energies of the Zr and Mo isotopes.}
 \end{figure}

In Fig.~\ref{fig:bind}, we plot calculated 
and experimental values of binding energies
with respect to the ${}^{88}\mathrm{Sr}$ core. We used 
the following formulae to obtain our binding energies:

\begin{eqnarray}
  \mathrm{BE}({}^{90+n}\mathrm{Zr}) & = & \mathrm{BE}({}^{90+n}\mathrm{Zr})-
  \mathrm{BE}({}^{88}\mathrm{Sr}) \nonumber \\
    & & -n \left[\mathrm{BE}({}^{89}\mathrm{Sr})-\mathrm{BE}({}^{88}\mathrm{Sr}) \right] \nonumber \\
    & & -2 
\left[\mathrm{BE}({}^{89}\mathrm{Y})-\mathrm{BE}({}^{88}\mathrm{Sr}) \right]
\end{eqnarray}

\begin{eqnarray}
  \mathrm{BE}({}^{92+n}\mathrm{Mo}) & = & \mathrm{BE}({}^{92+n}\mathrm{Mo})-
  \mathrm{BE}({}^{88}\mathrm{Sr}) \nonumber \\
    & & -n \left[\mathrm{BE}({}^{89}\mathrm{Sr})-\mathrm{BE}({}^{88}\mathrm{Sr}) \right] \nonumber \\
    & & -4 
\left[\mathrm{BE}({}^{89}\mathrm{Y})-\mathrm{BE}({}^{88}\mathrm{Sr}) \right]\;.
\end{eqnarray}

\noindent
An inspection of the resulting relative 
binding energies shows that the calculated values (shown by
asterisks) deviate from the experimental values 
(shown by filled circles), which display a parabolic
behavior.
This situation is common among calculations 
using realistic interactions derived from $NN$ scattering data \cite{holt98}.
It may be related to the absence of real three-body forces in the 
construction of the effective interactions \cite{zuker03}. 
It is known that a given Hamiltonian can always be separated in the form
$\hat{H}=\hat{H}_m+\hat{H}_M$, where $\hat{H}_m$ is the monopole part, while the
multipole $\hat{H}_M$ contains 
all other terms~\cite{abz1}. $\hat{H}_M$ given by realistic
$NN$ interactions takes proper account of the configuration 
mixing; however, the monopole part $\hat{H}_m$ fails to 
produce the correct 
unperturbed energies. It is possible to
change the averages of the so-called centroid 
matrix elements to fix this failure
without affecting the spectrosopy to produce the correct 
binding energies~\cite{duf1}.
However, a rigorous treatment of the global monopole corrections would deserve
a detailed study and thus  goes beyond the 
scope of the current work. Instead, to give some substance
to how a monopole correction may work, 
we add an overall constant to the diagonal interaction elements
so that the modified matrix elements are given by
\begin{eqnarray}
  V_J^{\mathrm{mod}}(ab,ab)=V_J(ab,ab)+ W\frac{n(n-1)}{2},
\end{eqnarray}
where $n$ is the 
number of valence particles. We have adopted $W=-125$ keV to reproduce
the binding energy of ${}^{102}\mathrm{Zr}$.
We plot the effect of this rather naive correction 
in Fig.~\ref{fig:bind}, where
modified results, represented by diamonds, show  
much better agreement for both chains of isotopes.

\subsubsection{$B(E2)$ strengths}
\begin{figure}
   \rotatebox{0}{\scalebox{0.62}{\includegraphics{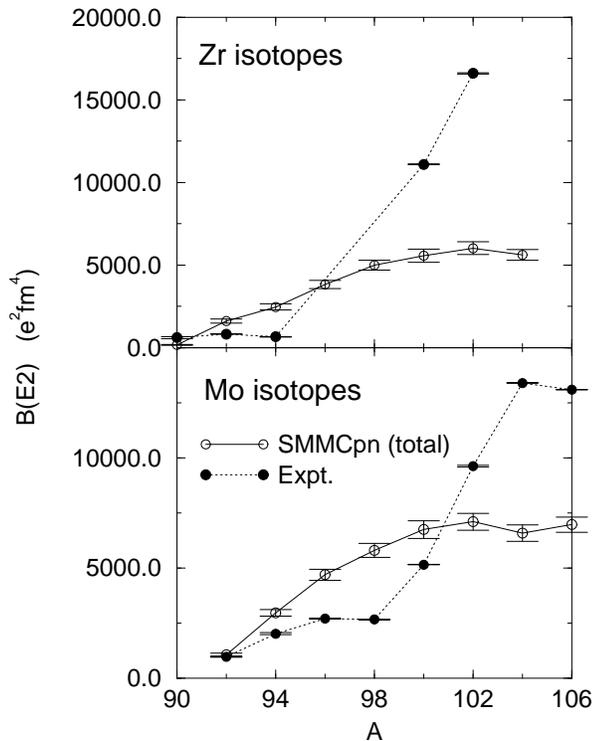}}}
 \caption{\label{fig:be2} Total $B(E2)$ strengths 
from the ground state for Zr and Mo isotopes.  Experimental results 
are for $0_1^+\rightarrow2_1^+$. }
 \end{figure}

Since the $2^+_1$ state is expected to 
absorb most of the total $B(E2)$ strength, the latter
can be used as a measure of 
the $0^+_1-2^+_1$ spacing, which should reflect a strong
change with the shape transitions.
Shown in Fig.~\ref{fig:be2} are the 
calculated total $B(E2)$ strengths (open circles) and
available experimental~\cite{ram1} $0^+_1 \rightarrow 2^+_1$ values
(filled circles). Despite the fact
that the calculated  total strengths increase as expected 
in both isotope chains with the
addition of neutrons,
their numerical values  fall somewhat below
the experimental 
$B(E2;0^+_1\rightarrow2^+_1)$ values on the heavier
side of the isotope chains. This failure is quite possibly 
due to our choice of single-particle energy for the 
$0h_{11/2}$ which is known to contribute to 
deformations in this region \cite{witek88}; it may also be due 
to correlations absent due to our choice of model space. We
will investigate this further in future work. 

\subsubsection{Pairing correlations}

\begin{figure}
   \rotatebox{0}{\scalebox{0.62}{\includegraphics{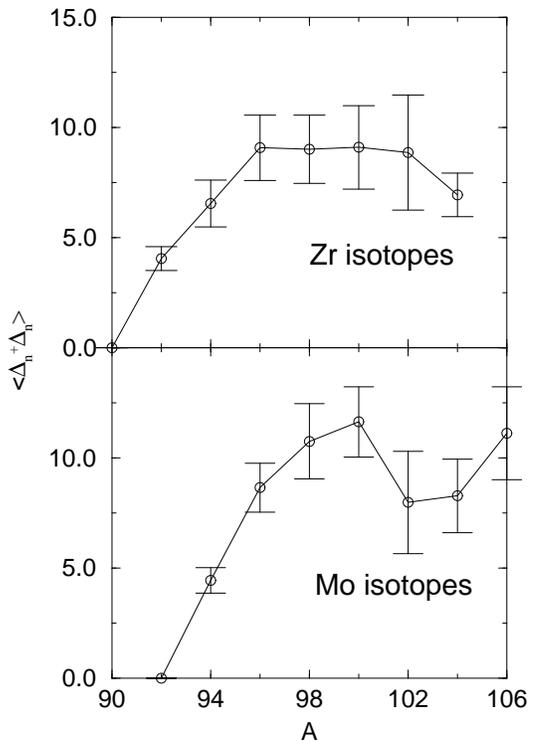}}}
 \caption{\label{fig:pairsn} Pairing correlations of Zr and Mo isotopes.}
 \end{figure}

Pairing correlations among like nucleons is 
known to be important for the ground state properties of
the even-even nuclei~\cite{dea1}. These correlations should become
quenched along the Zr and Mo isotope chains as the transition from spherical to
well-deformed shapes becomes more pronounced. Similar effects were recently
investigated in $N=40$ isotones \cite{langanke05}. 
We have investigated the pairing content of the 
ground states of the nuclei of interest using
a BCS-like pair operator which is  defined for neutrons as

\begin{equation}
  \hat{\Delta}^{\dagger}_{\nu}=\sum_{jm>0} \nu_{jm}^{\dagger} \nu_{j\bar{m}}^{\dagger},
\end{equation}

\noindent where the sum is over all orbitals  with $m>0$, and
$\nu_{j\bar{m}}^{\dagger} = (-1)^{j+m}\nu_{j-m}^{\dagger}$ is the time-reversed operator.
Hence the expectation value of the pairing fields is
$\langle \hat{\Delta}^{\dagger}_{\nu} \hat{\Delta}_{\nu} \rangle$.
This quantity for an uncorrelated Fermi gas is given by
\begin{equation}
  \langle \hat{\Delta}^{\dagger} \hat{\Delta} \rangle = \sum_j \frac{n_j^2}{2(2j+1)},
\end{equation}
where $n_j=\langle \nu_{jm}^{\dagger} \nu_{jm} \rangle$ are the 
neutron occupation numbers.
Any excess over the Fermi-gas value 
therefore indicates pairing correlations in the ground state.
Our results, which are plotted in 
Fig.~\ref{fig:pairsn}, confirm a suppression of these correlations,
as the contribution of the added 
neutrons to the pairing gradually decreases and the correlations
become noticeably
quenched beyond ${}^{96}\mathrm{Zr}$ and ${}^{100}\mathrm{Mo}$.

\begin{figure}
   \rotatebox{0}{\scalebox{0.45}{\includegraphics{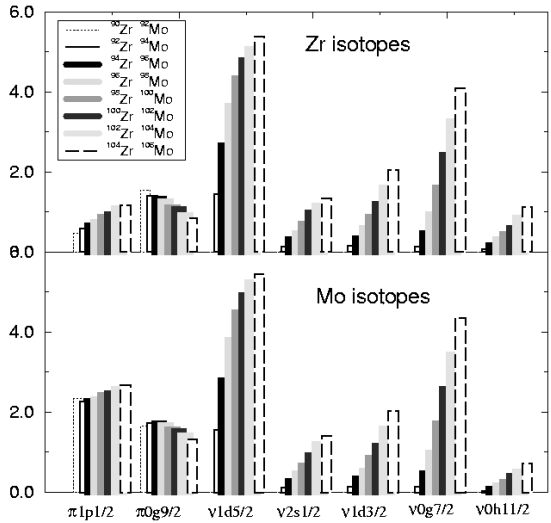}}}
 \caption{\label{fig:rho} Orbital occupation numbers.}
 \end{figure}

In addition, the occupation numbers of various 
orbitals are plotted in Fig.~\ref{fig:rho},
demonstrating that additional
neutrons are distributed into the available orbitals rather uniformly,
while  protons tend to migrate from the $0g_{9/2}$ to the $1p_{1/2}$ orbital.
In the case of Mo isotopes, the 
spurious effect of exceeding the maximum-allowed occupancy
for the $1p_{1/2}$ orbital is a result of 
the extrapolation scheme that was used, and is indicative that the 
relative error bars on the occupation data after extrapolation are 
approximately 15\% of the value of the occupation number.  Note that the
deformation driving $0h_{11/2}$ remains only slightly occuppied. 

\section{Conclusion and perspectives}

We have introduced a new approach for the implementation of the
Shell Model Monte Carlo method to perform shell-model
calculations using non-identical proton and neutron valence spaces.
General features of the SMMC method were reviewed. 
Differences between the
isospin and the $pn$-formalisms have been 
pointed out; in particular,
the $T_z$ projection has been described in detail.

The results of the SMMCpn approach 
have been validated in the $sd$-shell using a ``good"-signed
schematic interaction and in 
the $fp$-shell  using the realistic KB3 interaction.
In the latter case, we dealt with the sign 
problem using an extrapolation method. 

As the first novel application of the SMMCpn
approach, we performed a set of calculations
for the even-even ${}^{90-104}\mathrm{Zr}$ and ${}^{92-106}\mathrm{Mo}$ 
isotopes, using a realistic effective interaction in the 
valence space described by
($\pi\!:$ $1p_{1/2},0g_{9/2}$) and 
($\nu\!:$ $1d_{5/2},2s_{1/2},1d_{3/2},0g_{7/2},0h_{11/2}$)
orbitals. A comparison of the ground-state 
energies of the first few nuclei in both isotope chains
showed  excellent agreement with the exact diagonalization results and 
provides a definitive test of our algorithm and the SMMCpn method. 
We then studied the transitional nature 
of the isotopes by using the  $B(E2)$
strength as a gross measure of the $0_1^+-2_1^+$ separation.
Along both isotope chains, we have obtained an 
enhancement in the $B(E2)$ strengths as a function of
the added neutrons, accompanied by a 
quenching in the neutron-pairing correlations.
In spite of  this qualitative reproduction of the onset of deformations,
further research clearly is needed for a qualitative result in the
heavier Zr and Mo nuclei. 
A comparison with the experimental data suggests that
this situation may be a shortcoming due to the degrees of
freedom that are absent in the chosen valence space and possibly due to 
the value of the $0h_{11/2}$ orbital. We will investigate this further
in future work. 

Apart from future applications involving realistic 
effective interactions, use of
schematic interactions in SMMCpn applications should 
be an interesting direction of research.
Such interactions have been commonly used 
to calculate realistic estimates of
collective properties and level densities; 
the latter is an important ingredient in the prediction
of nuclear reaction rates in astrophysics. Parity 
dependence of these densities may play a crucial role in the nucleosynthesis.
We believe that SMMCpn will prove to be 
a useful computational tool in this regard.

\begin{acknowledgments}
Research was supported by the U.S. Department of Energy under
Contract Nos. DE-FG02-96ER40963 (University of Tennessee),
and DE-AC05-00OR22725 with UT-Battelle, LLC (Oak Ridge
National Laboratory). We acknowledge 
useful discussions with M. Hjorth-Jensen. 
\end{acknowledgments}


\end{document}